\documentclass[journal]{IEEEtran}

\ifCLASSINFOpdf
\else
\fi
\usepackage{algorithm}
\usepackage{algpseudocode}
\usepackage{amsmath}
\usepackage{graphicx} %use graph format
\usepackage{epstopdf}
\usepackage{subfigure}
\usepackage{cite}
\usepackage{multirow}
\usepackage{booktabs}
\usepackage{color}
\newtheorem{defn}{Definition}
\newtheorem{theorem}{Theorem}

\begin{document}

\title{Cost Restrained Hybrid Attacks in Power Grids}

\author{Xiaolin~Gao, Cunlai~Pu, and Lunbo~Li
\thanks{X. Gao, C. Pu and L. Li are with the School of Computer Science and Engineering, Nanjing University of Science and Technology, Nanjing 210094, China. Corresponding author: C. Pu; Email: pucunlai@njust.edu.cn. }}
\maketitle

\begin{abstract}
The frequent occurrences of cascading failures in power grids have been receiving continuous attention in recent years. An urgent task for us is to understand the cascading failure vulnerability of power grids against various kinds of attacks. We consider a cost restrained hybrid attack problem in power grids, in which both nodes and links are targeted with a limited total attack cost. We propose an  attack centrality metric for a component (node or link) based on the consequence and cost of the removal of  the component. Depending on the width of cascading failures considered, the attack centrality can be a local or global attack centrality.  With the attack centrality, we further provide a greedy hybrid attack, and an optimal hybrid attack  with the Particle Swarm Optimization (PSO) framework.
 Simulation results on IEEE bus test data show that the optimal  hybrid attack is  more efficient than the greedy hybrid attack. Furthermore, we find counterintuitively that the local  centrality based algorithms are better than the global centrality based ones when the cost constraint is considered in the attack problem.
% A comparison is made with two other greedy attack strategies: the general attack centrality based hybrid attack (GCHA) and the random hybrid attack (RHA). Simulation results show that the optimized hybrid attack can achieve higher performance than the greedy attacks. Among all the attack strategies, the local centrality based hybrid attack (LCHA) has the lowest computational cost.
Our work can help in the robustness optimization of power systems  by revealing  their worst-case attack vulnerability and  most vulnerable components.
\end{abstract}


\IEEEpeerreviewmaketitle

\section{Introduction}
\IEEEPARstart{I}{n} power grids, a small-scale broken down of components can result in large-scale cascading failures or blackouts that cause catastrophic consequences. In the past year, large blackouts have hit many places in the world including the US, Australia, Argentina, and Venezuela. The scale and frequency of blackouts are continuously growing due to the increasing interconnection in power infrastructures. The causes of the cascading failures can be the faults at power stations, damage to electric transmission lines, and deliberate network attacks. It is of great significance to have a profound understanding of the cascading failure vulnerability of power girds.

\par The topic of cascading failures has received great attention from the area of network science. A branch of research work was focused on the modeling and analysis of cascading failures \cite{xia2019introduction}. Ren et al. \cite{ren2018stochastic} proposed a cascading failure model for communication networks, in which each overloaded element has a probability of failure.  Jiang et al.  \cite{jiang2018cascade} provided a heuristic algorithm to discover key nodes in cascading failures. Chen et al. \cite{chen2017robustness} further performed a critical node analysis on interdependent power  and communication networks to identify the vital nodes in cascading failures. Lee et al. \cite{lee2019vulnerability} considered the data injection attacks and provided the associated  load redistribution model in cascading failures.  Pu et al. \cite{pu2019vulnerability} proposed a link centrality measure based on both topological and electrical properties and further studied the cascading failures induced by the link centrality based attacks.

 On the other hand, various works have been devoted to the defense of cascading failures. Jiang et al. \cite{jiang2017efficient} proposed a local load redistribution mechanism with the given node capacity distribution, which is  applied to the mitigation of cascading failures.   Zhou et al. \cite{zhou2018network} investigated  the impending breakdown prediction of the network by monitoring the critical indicators, and further provided a node addition strategy to prevent the collapse of the network under cascading failures.  In order to restore the network during cascading failures, Huang et al. \cite{huang2018sequential} gave the result-oriented and resource-oriented restoration approaches respectively. Xu et al. \cite{xu2020effect} discussed the optimization of the allocation of the limited recovery resources  to repair the failed nodes in cascading failures. Tu et al. \cite{tu2018optimal} studied how the topological metrics affect the robustness of power grids and used the simulated annealing method to find the optimal network topology against cascading failures.  Wang et al. \cite{wang2017enhancing} discussed three optimization algorithms aiming at enhancing network robustness against cascading failures included by link attacks.

\par As far as we know, the existing work   considered attacking either nodes or links that the hybrid attack including both  types of components has been rarely reported, which, however,  happens in real situation.  More importantly, an attack usually has a cost in any situation, and the costs for attacking a node and a link are usually different.  Since removing  a node is equivalent  to  removing all the links associated with the node, it is reasonable to assume that the cost for attacking this node is equal to the cost for attacking all the links incident from the node.  Therefore, from an adversary's perspective,  it needs to optimize the selection of nodes and links for attacks when the total attack cost is given, in order to achieve a large attack performance.
 %Ye et al. \cite{ye2016optimal} investigated  a cost constrained attacks on networks, in which they only consider the cost of attacking nodes.  A challenging problem which arises in this domain is to study the hybrid attacks.

\par In this paper, we study the robustness of power grids against the cost restrained hybrid attacks   and their resulting cascading failures. The contributions of our work are summarized as follows.

\begin{itemize}
  \item We formalize a cost restrained hybrid attack problem. In particular,  we seek the best  combination of nodes and links in power grids with a given total attack cost, removing which will lead to the largest scale of cascading failure.  We prove that this optimization problem is NP-hard.
  \item Based on the damage and cost of attacking a component, we propose an attack centrality metric. With this metric, we further provide  greedy hybrid attack algorithms and  optimal hybrid attack algorithms, which embody the  particle swarm optimization (PSO) framework.
\end{itemize}

\par The rest of this paper is organized as follows. We provide the cascading failure model in Section \uppercase\expandafter{\romannumeral2}. In Section \uppercase\expandafter{\romannumeral3}, we present the cost restrained  hybrid attack problem. In Section \uppercase\expandafter{\romannumeral4},  we introduce our  attack centrality measurement and its related greedy  and optimal attacks.  Simulation results are presented in Section \uppercase\expandafter{\romannumeral5}. Finally, we conclude our work  in Section \uppercase\expandafter{\romannumeral6}.

\section{Model of Cascading Failures}
To obtain the topological structure of a power grid, we can take the  base stations and transmission lines simply as  nodes and links respectively, and then the corresponding network topology can be represented by an undirected graph $G(V,E)$, where $V$ and $E$ denote the sets of nodes and links, respectively. The grid network has then  $N=\left|V\right|$ nodes and  $M=\left|E\right|$ links accordingly.

Following Ref. \cite{tu2018optimal}, we consider two types of nodes:  generation nodes and  consumer nodes. Then, the corresponding admittance matrix of a power grid can be written as

\begin{equation}
\left[
\begin{array}{cccc}
\dots&\dots&\dots&\dots\\
\dots& y_{i} & 0 &\dots\\
\dots& -Y_{ji} &Y_{jj}&\dots\\
\dots&\dots&\dots&\dots
\end{array}
\right ]
\left[
\begin{array}{cccc}
\dots\\v_{i}\\v_{j}\\\dots
\end{array}
\right ]
=
\left[
\begin{array}{cccc}
\dots\\v_{i}\\I_{j}\\\dots
\end{array}
\right ],
\end{equation}
which is obtained according to the Kirchhoff's law equations.
In Eq. (1), the subscripts $i$ and $j$ represent generation node and consumer node, respectively; $v$ and $I$ represent the voltage and external injected current, respectively; the element $Y_{ij}$ is the admittance of the link connecting nodes $i$ and $j$, and $Y_{ij}=0$ if  nodes $i$ and $j$ are not connected.  Moreover, we have {$y_{i}=1$ for a generation node $i$ and $Y_{jj}=-\sum_{s\neq{j}}Y_{js}$ for a consumer node $j$}, which  are ensured  by the Kirchhoff's law. Usually, the grid topology, link admittances,  voltages of generation nodes, and  current consumptions of consumer nodes are given as the prior information.  Then, the voltages of consumer nodes can be obtained using Eq. (1).

Following Ref \cite{tu2018optimal}, we define that the load on link $\left(s,d\right)$ is equal to  its current flow, $I_{sd}=\left( v_{s} - v_{d} \right)\times Y_{sd}$; the load on node $s$ is  $v_{s} \times I_{os}$, where $I_{os}$ is the total current flowing out of node $s$. The capacity (maximum allowed load) of a node   is defined as  $(1+\alpha)$ times  its initial load, and the capacity of a link is set to be $(1+\beta)$ times  its initial load, where  $\alpha$ and $\beta$ are safety margins of nodes and links, respectively. If the load of a component exceeds the given capacity, the component is considered to be  failed and will be removed from the grid.

\par In power grid, an initial small-scale failure of components can further cause the overload of other components and eventually the cascading failures.  The whole process of a cascading failure in the power grid, can be modeled as follows:
\begin{enumerate}
    \item Initially, some components are removed from the grid, which usually represents the initial attack.
  \item The  load of each remaining component  is recalculated, and if the current load of a component exceeds the given capacity, it will be removed immediately  from the grid.
  \item It is possible that the grid are split into multiple isolated subgrids after the removal of components.  For a subgrid containing no generator nodes, all components in this subgrid are considered as unserved and are removed accordingly.
  \item Repeat steps 2) and 3) until all existing components operate under capacity constraints.
\end{enumerate}
\par A simple way for quantifying the consequence or damage of an attack on the grid is to calculate the fraction of the failed components. Usually, we exclude the number of components manually removed at the beginning to fairly evaluate the performance of an attack.  Based on this idea, the damage of an attack is given by

\begin{equation}
\Phi=\frac{N_{unserved}-N_{attacked}}{D-N_{attacked}},
\end{equation}
where $N_{unserved}$ is the total number of unserved components after cascading failures; $N_{attacked}$ is the number of initially removed  components;  $D$ is the number of all components at the beginning,  which equals to $N+M$.

\section{Problem Formulation}
In our cost constraint hybrid attack model, we remove a selected set of components containing both nodes and links from the grid under a cost constraint, which acknowledges the fact that usually removing  a  node or link  has a cost. Generally, the larger the admittance of a circuit,  the larger the price of the circuit is \cite{smith1982microelectronic}. Thus, we  assume the cost of removing a link is $\gamma$ times its admittance. Since the removal of a node is equivalent to the removal of all its incident links, we further assume that the attack cost of a  node is the sum of the cost of all the links incident on the node. For component $i$, we denote its attack cost  by $c_i$. The maximum total attack cost allowed in the attack  is given   to be proportional to the sum of the attack cost of all components in the grid,
\begin{equation}
C_0=\theta\sum_{i=1}^{D}c_{i},
\end{equation}
where $\theta\in[0,1]$ is the control parameter of total attack cost $C_0$. The larger $\theta$ generally means more components can be removed in the attack.

\par We define the solution of an attack as  $X=\left[x_1,x_2,\cdots,x_D\right]$, where $x_i=1$ if component $i$ is attacked, otherwise $x_i=0$. The cost of the attack solution $X$ can be written as
\begin{equation}
C(X)=\sum_{i=1}^{D}c_i*x_i.
\end{equation}
\begin{defn}[Optimal Component Set Problem]
 Given an attack cost $C_0$,  find the optimal component set, the removal of which  will result in the largest damage  $\Phi$.
\end{defn}
\par The optimal component set problem (OCSP) can be formalized as
\begin{equation}
\begin{aligned}
& \underset{X}{\text{maximize}}
&&\Phi(X)\\
& \text{subject to}
&& C(X)\leq C_0.
\end{aligned}
\end{equation}
\par Note that the number of attacked components $K$ is not given as precondition, which eventually depends on the optimal attack solution.
\begin{theorem}
The $OCSP$ is $NP\textrm{-}hard$.
\end{theorem}

\begin{IEEEproof}
%First, due to the optimization form of the $OCSP$, we cannot achieve the optimal solution in polynomial time, or given an attack, there is no polynomial algorithms to check whether the solution is the optimal, which is similar to the optimization problem form of the 0/1 knapsack problem \cite{Toth1990Knapsack}. It means that the $OCSP$ is not in $NP$.
For a meaningful attack, the number of failed components in the cascading failure should be significantly larger than the number of components manually removed at the beginning, i.e., $N_{unserved}\gg N_{attacked}$. Thus, the damage $\Phi$ is  approximately linearly dependent on $N_{unserved}$ based on Eq. (2).
%Given an attack set $S$ with size $\left|S\right|$ for the $OCSP$, the $PUC$ caused by attacking $S$ can be written as $\Phi(S)=\frac{N_{unserved}^{c}-\left|S\right|}{D-\left|S\right|}$. Suppose that there is a non-negative number $k$ satisfying $\left|S\right|\geq k$ (this is obviously easy to achieve), then there is $\Phi(S)\geq \frac{N_{unserved}^{c}-k}{D-k}$. Since $D$ and $k$ are both constants, it can be roughly considered that $\Phi(S)$ is large enough as $N_{unserved}^{c}$ is large enough.
In other words,  the $OCSP$ is essentially  equivalent to seeking an optimal constrained hybrid attack that maximizes $N_{unserved}$. On the other hand, the problem of maximizing $N_{unserved}$ is similar with the 0/1 knapsack problem \cite{Toth1990Knapsack}: given $N$ items numbered 1 to $N$, each with a weight $w_i$ and a value $v_i$, seek a collection $S$ of items such that $\sum_{i\in S} w_i \leq W$ (the total weight constraint) and the total value $\sum_{i\in S} v_i$ is as large as possible. In the problem of  maximizing $N_{unserved}$, if we set $D=N$,  $c_i=w_i$ and  $N_{unserved}(i)=v_i$ ($N_{unserved}(i)$ represents the number of unserved components caused by the removal of component $i$), then this problem is equivalent to the 0/1 knapsack problem. Note that we can artificially fix the  double-counting of failed components in the calculation of $N_{unserved}(i)$.  Collectively,  the 0/1 knapsack problem can be reduced to the $OCSP$ in polynomial time.
  %the  Once given a set $S$, if the cascading failure caused by attacking $S$ is non-repetitive, the final profit can be calculated by $\sum_{i\in S} N_{i}^{uc}$ so that the problem to maximize $N_{unserved}^{c}$ is equivalent to the 0/1 knapsack problem. In fact, we can use digital processing to make the range of failures of attacking an independent component in $S$ non-repetitive, and accordingly a solution for the 0/1 knapsack problem also satisfies the problem of seeking the maximal $N_{unserved}^{c}$ and corresponding $\Phi$. For the reverse direction, given an attack which maximizes $\Phi$ and accordingly a satisfying $N_{unserved}^{c}$, the set constitute a solution to the 0/1 knapsack problem.
Since the 0/1 knapsack problem is a well known $NP\textrm{-}hard$ problem,  the $OCSP$ is also $NP\textrm{-}hard$.
\end{IEEEproof}

%Due to the NP-hardness, it is in general lacking of systematic methodology to solve the optimization problem above efficiently.

\section{Attack centrality  and its applications in algorithm design}
In this section, we first propose an attack centrality metric, which considers both the damage and cost of removing a component. Then, we employ the attack centrality  to design  greedy and optimal algorithms. We  also discuss the computational cost of the proposed  algorithms.

\subsection{Attack centrality }
Different from previous work, where the attack consequence was the only concern,  we consider  the attack consequence and attack cost simultaneously in our work. Specifically, we define the attack centrality of a component as the ratio of the total attack cost of the failed components caused by the removal of the component to the attack cost of the component itself. For component $i$, its attack centrality is
\begin{equation}
\psi(i)=\frac{\sum_{j \in \Omega_i} c_j}{c_i},
\end{equation}
where $\Omega_i$ is the set of components failed in the cascading failures triggered by the removal of component $i$. This centrality metric essentially  indicates the cost-effectiveness of attacking a single component. Note that in the simulation,  the computational cost increases with the increase of the width of the cascading failure, and the total cascading width may be different for different components. Thus, we consider  two special cases of cascading width, i.e., width one and the total width of a cascading failure, corresponding to the local and global attack centralities, respectively. In other words, to measure the local centrality of a component, we only consider the failed components of width one in the cascading failure caused by the removal of the component, while considering  all failed components in the whole cascading process to calculate its global centrality.

\subsection{Greedy hybrid attack}
In the greedy attack, we first rank all components in the decreasing order of their attack centrality. Then, we select the components from the top of the ranking under the given constraint of the  total attack cost. Finally, we remove the selected components simultaneously from the grid, and evaluate the subsequent cascading failure based on Eq. (2).

The proposed local and global attack centralities  are employed in the greedy attack and thus yield two greedy attack algorithms, which are  called local centrality based greedy hybrid attack (LC-GHA) and global centrality based greedy hybrid attack (GC-GHA), respectively. We also consider the random hybrid attack (RHA) for the comparative purpose, in which  we randomly remove a set of components under  the total attack cost limit.

The time complexity of the greedy hybrid attack is $O(D)$, when the width of a cascading failure is ignored due to its randomness. In practice,  LC-GHA  always has a much lower computational cost than GC-GHA, since only width one of a cascading failure is considered in the calculation of attack centrality for  LC-GHA.     Also,  RHA  has the lowest computational complexity, since it does not need to  calculate attack centrality.

\subsection{Optimal hybrid attack}
Since the OCSP is  NP hard, we use the particle warm optimization (PSO)  framework to seek its optimal solution.
The general procedure of the PSO in our attack scenario is  as follows.
\begin{enumerate}
  \item Initialize $n$ particles. A particle $i$ is represented by  a $D$-dimensional vector  $X_i$, in which each element is independently set to a random value in [0, 1]. Let $\widetilde{X}_i$ be a replica of  $X_i$. We then set  the largest $K$ elements in $\widetilde{X}$ under the total attack cost constraint to 1 and the rest to 0.
  \item Calculate  $\Phi(\widetilde{X}_i)$ by using Eq. (2) for each particle $i$, which is also called the fitness in PSO. Then, update the local and global optimal solutions for the particle swarm.
  \item Update the velocity and position of the particles. Repeat step 2) until the maximum  number of iterations $t_{max}$ is reached.
\end{enumerate}

%\par The final solution carried out is the optimal $K$-component set at $iter_{max}$.

 The velocity and position of particle $i$  update  with the following equations:
 \begin{equation}
 \hspace{-0.12cm}
\begin{cases}
\hspace{-0.05cm}{v_i}^{t+1}\hspace{-0.1cm}=\hspace{-0.1cm}w^{t}{v_i}^t\hspace{-0.1cm}+\hspace{-0.1cm} c_1{r_1}^t(p_{best}^i\hspace{-0.1cm}-\hspace{-0.1cm}{x_i}^t)\hspace{-0.1cm}+\hspace{-0.1cm}c_2{r_2}^t(g_{best}\hspace{-0.1cm}-\hspace{-0.1cm}{x_i}^t),\\
\hspace{-0.05cm}{x_i}^{t+1}\hspace{-0.1cm}=\hspace{-0.1cm}{x_i}^t\hspace{-0.1cm}+\hspace{-0.1cm}{v_i}^t,\\
\hspace{-0.05cm}w^{t+1}\hspace{-0.1cm}=\hspace{-0.1cm}w_{max}\hspace{-0.1cm}-\hspace{-0.1cm}\frac{t*(w_{max}-w_{min})}{t_{max}}.
\end{cases}
\end{equation}
\par In the above equations, $v_i^{t}$ is the velocity of node $i$ at time step $t$; $x_i^{t}$ represents the position of node $i$ at time step $t$;  $w^{t}$ is the inertia coefficient at time step $t$; $w_{max}$ and $w_{min}$ are respectively the maximum and minimum values of $w$; $c_1$ and $c_2$ are respectively the cognitive coefficient and the social learning coefficient; both $r_1^{t}$ and $r_2^{t}$ are random values  in~[0, 1], which are independently generated at time step $t$.
 %For the sake of simplicity, we neglect the effect of parameters on the results.

\par In theory, the PSO based optimal hybrid attack  algorithm, named as OHA,   can find out  the optimal attack solution $X_{opt}$ under a given total attack cost constraint   for a sufficiently large number of iterations. Nevertheless, in real situations  $t_{max}$ is always limited, and the quality of the solution is not guaranteed. Therefore, we further improve the OHA by applying our attack centrality to the optimization of the initial solution.  Specifically, in the initialization phase, we randomly select  one particle and set the corresponding $X$  with the values of the attack centrality of the components.
%In order to improve the performance and efficiency of the optimization algorithm, we propose a local attack centrality based optimal hybrid attack strategy $(LCOHA)$ according to the $GOHA$ presented above. In the initialization phase, we correspond one $\vec{X_{copy}}$ of the particles to the ranking according to the local attack centrality so that the initial solution is relatively better.
 Moreover, inspired by the genetic algorithm \cite{whitley1994a}, we utilize the  mutation mechanism to reset the particles  to avoid their falling into the local optima. Specifically, after each iteration,  the mutation from 0 to 1 or 1 to 0 is carried out for  an  element of the current optimal solution, which is selected randomly.
 Depending on the local or global attack centrality used in the optimization, we have local centrality based optimal hybrid attack (LC-OHA) or global centrality based  optimal hybrid attack (GC-OHA), respectively. The time complexity of these two optimal hybrid attack algorithms are both $O(n*t_{max})$ without considering the duration of the single cascading failure process.
 The pseudocode of the centrality based optimal hybrid attack is in Algorithm 1.
\begin{algorithm}
  \caption{Attack centrality based optimal hybrid attack}
  \begin{algorithmic}[1]
    \Require
       The given total attack cost  $C_0$; Number of particles $n$; Particle dimension $D$; Number of iterations $t_{max}$;
    \Ensure
      Optimal $K$-component set;
    \State Calculate the  attack centrality of each component;
    \State Initialisation: Given $n$ particles, each one is represented by a $D$-dimensional vector $X$. For particle $i$, independently set each  element of $X_i$ to be a random value in [0, 1]. Randomly select a particle $X$, and reset its elements to be the values of the attack centrality of the components, i.e., $x_k=\psi(k)$.  Assume $\widetilde{X}_i$ is a replica of $X_i$, and then set the largest $K$ elements in $\widetilde{X}_i$ to 1 and the rest to 0 with the constraint $C(\widetilde{X}_i) \leq C_0$;
     %Randomly select a  $\widetilde{X}$ and set its $K$ elements, which have  the largest attack centrality values and satisfy the total cost constraint,  to 1 and the rest to 0;
    \For {each particle}
      \State Calculate its fitness $\Phi(\widetilde{X})$;
    \EndFor
    \State Update the global optimal value $g_{best}$ of the particle swarm;
    \For {$t=1 : t_{max}$}
      \State Generate a random number $b \in[0, 1]$;
      \For {$j=1 : n$}
        \If {$j==b$}
          \State $X_j \leftarrow \ g_{best}$;
          %\For {$p=1:b$}
            \State Select an element of $X_j$ and conduct the mutation from  0 to 1 or 1 to 0;
          %\EndFor
        \EndIf
        \State Update the velocity and position of particle $X_j$;
        \State Copy $X_j$ to $\widetilde{X}_j$ and set the largest $ K$ elements in $\widetilde{X}_j$ to 1 and the rest to 0 under the cost limit;
        \State Calculate the fitness $\Phi(\widetilde{X}_j)$;
        \State Update the optimal solution $p^j_{best}$ of particle $j$ and the global optimal solution $g_{best}$ of the particle swarm;
      \EndFor
    \EndFor
    \State \Return the optimal $K$-component set corresponding to $g_{best}$.
  \end{algorithmic}
\end{algorithm}

%The time complexity of the two optimized hybrid attack algorithms are both $O(n*iter_{max})$ without considering the duration of the single cascading failure process. Note that the cost restrained model in the optimization algorithms is in keeping with the greedy hybrid attack strategies proposed above.\setlength{\abovedisplayskip}{3pt}
\section{SIMULATION RESULTS}
In this section, we evaluate the performance of the proposed algorithms on IEEE 118 bus and IEEE 162 bus data by using MATLAB. The topologies of the networks and  the generation nodes are all given in the data.   For the simplification purpose, we  set that the voltage of generation nodes is 1 p.u. (per unit), the current of the consumer nodes is 1 p.u., and the admittances of the links  follow a normal distribution with an average value of 11 p.u. and a standard deviation of 2 p.u, which  basically follow  ref. \cite{tu2018optimal}.  The cost parameter $\gamma$ is set to be 0.3. The safety margins are usually small and thus set to be $\alpha=\beta=0.2$ in the experiments. The  parameters
for the particle swarm are empirically set as  $n=10$, $c_1=c_2=2$, $w_{max}=0.9$, $w_{min}=0.4$ and $t_{max}=200$.
\begin{figure}
  \setlength{\abovecaptionskip}{0.cm}
  \setlength{\belowcaptionskip}{-0.2cm}
  \centering
  \hspace{-10mm}
  \subfigure[IEEE-118-bus]{
    \includegraphics[width=0.275\textwidth,height=3.8cm]{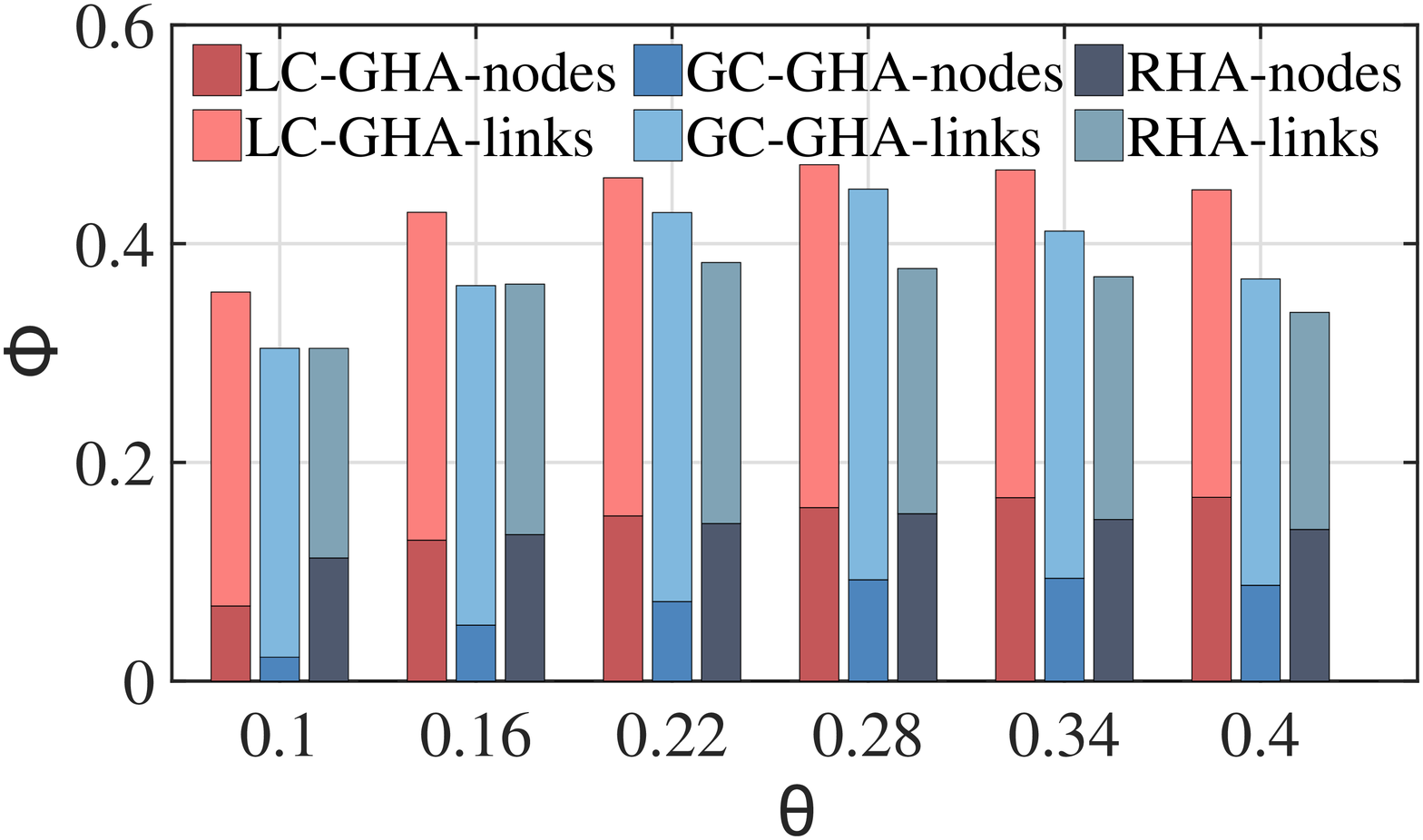}}
    \hspace{-7mm}
  \subfigure[IEEE-162-bus]{
    \includegraphics[width=0.275\textwidth,height=3.8cm]{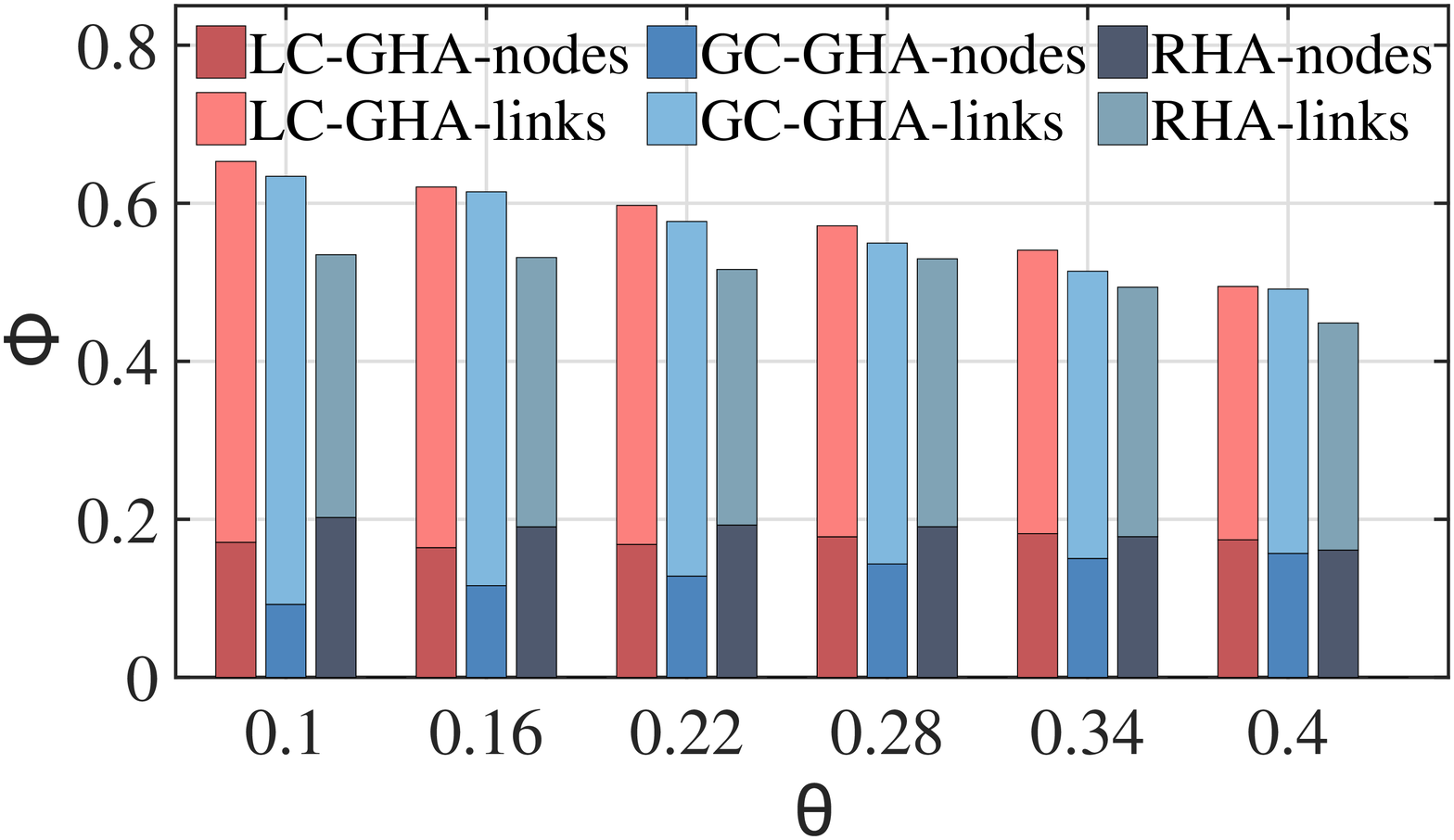}}
    \hspace{-10mm}
  \caption{The damage $\Phi$ vs. the cost control parameter $\theta$ under three greedy hybrid attack strategies: LC-GHA, GC-GHA and RHA on two IEEE bus test data.}
  \vspace{-0.4cm}
\end{figure}
\par First, we present the results of the greedy hybrid attacks, i.e.,   LC-GHA and GC-GHA, with a comparison with RHA in Fig. 1, which are the average of 50 independent runs. For each histogram, the light and dark colors correspond to the fractions of links and nodes among all the attacked components, respectively. We see from Fig. 1 that instead of single type of components, both links and nodes are attacked for all the three  algorithms to achieve a good attack performance. This indicates that when attack cost is also a major concern, it is necessary to weigh all types of components for a cost-effective  attack.

 Furthermore, the damage $\Phi$ of LC-GHA  is larger than the GC-GHA and RHA  for the same attack cost.
% The RHA  almost always has the worst performance among the three due to nature of random selection.
  The evidence that LC-GHA is better than GC-GHA is quite counterintuitive, since the global centrality calculated with more information and computational cost is expected to be more effective  than local centrality in a general sense.
 In our cost constrained hybrid attack model, the attack centrality of a component is a trade-off between the attack performance and attack cost. This yields that a node with a larger attack performance may not has a larger attack centrality than a link with a smaller attack performance, since node generally has a larger attack cost than link. Therefore, the links with large attack performance especially in the GC-GHA are attacked with high priority, which is why the fraction of attacked links is always larger than that of attacked nodes in the greedy attacks. The larger tendency to links in the GC-GHA compared to that in the LC-GHA leads to the lower efficiency of the former than the latter. RHA always has the lowest efficiency due to  its nature of randomness.

\begin{figure}[htbp]
  \centering
  \setlength{\abovecaptionskip}{-0.2cm}
  \includegraphics[width=0.48\textwidth,height=4.5cm]{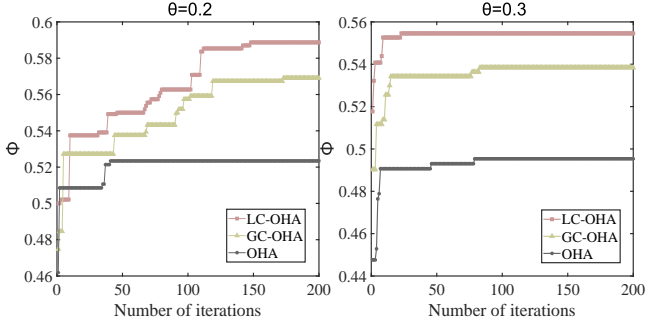}
  \caption{ The damage $\Phi$ vs. the number of iterations for LC-OHA, GC-OHA and OHA on IEEE 118 bus data with $\theta=0.2$ and $\theta=0.3$, respectively. }
  %The light and dark colors of the subgraphs show the number of links and nodes targeted at $t_{max}$.}
  \vspace{-0.4cm}
\end{figure}

\par Next, we evaluate the performance of optimal hybrid attacks, i.e., LC-OHA and GC-OHA, with a comparison with OHA. Fig. 2 shows the relation between the damage $\Phi$ and the number of iterations for the three optimal hybrid attacks.  We can see clearly that with the increase of the iterations, the damage generally increases and then converges for all the optimal hybrid attacks. Interestingly, LC-OHA converges faster with a higher damage than GC-OHA, which together with its lower computational cost demonstrates its advantage over  GC-OHA. The OHA always has the lowest efficiency owing to no optimization of initialization.

\begin{figure}
  \centering
  \setlength{\abovecaptionskip}{-0.2cm}
  \includegraphics[width=0.36\textwidth,height=4.8cm]{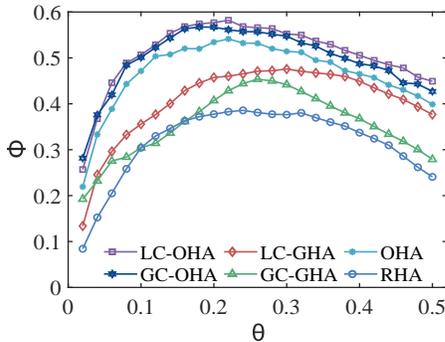}
  \caption{ The damage $\Phi$ vs. the cost control parameter $\theta$ for all the greedy and optimal hybrid attacks on IEEE 118 bus data.}
  \vspace{-0.4cm}
\end{figure}

\par  Finally,  we compare  the performance of all the greedy and optimal hybrid attacks. The results of damage vs. cost constraint are given in Fig. 3, which are the average of 100 independent realizations.   We see that the  optimal hybrid attacks noticeably outperform the  greedy hybrid attacks for achieving a larger attack performance under an arbitrary  cost constraint. The decreasing order of efficiency is LC-OHA~$>$~GC-OHA~$>$~OHA~$>$~LC-GHA~$>$~GC-GHA~$>$~ RHA, which further confirms  that the local attack centrality is better than the global one in the cost-constrained hybrid attack problem. Note that generally the greedy attacks have a much smaller computational cost than the optimal attacks.  In addition, we observe that with the relaxation of cost limit, the damage $\Phi$ first increases and then decreases after the peak for all hybrid attacks.
 %and the peaks of optimal attacks come earlier than the greedy ones.
  When the allowed total attack cost is small, the damage increases with the cost, i.e., the scale of failed components in the cascading failure increases with the growth of the attack cost. While when the cost constraint is large enough,  the number of failed components (except those removed at the beginning) tends to be stable as the attack cost grows, which accounts for the decrease of $\Phi$ (see Eq. (2)).
%   Assuming that the stable value of the damage scale caused by attacking a set $S$ is $K$ when the cost limit is large, the current $\Phi=\frac{K-\left|S\right|}{D-\left|S\right|}$, which is a function with $\left|S\right|$ as its variable. The trajectory of the curve to the right of the peak in Fig. 3 validates our conclusion. Note that the greedy algorithms  have a much smaller computational cost than the optimal algorithms.

\section{Conclusion}
In summary, we have first discussed the  cost constrained hybrid attacks and their induced cascading failures in power grids. The formulated   optimization problem, which is proved to be NP hard,  aims at finding the optimal set of components to be attacked under a given attack cost constraint. We proposed an attack centrality measure by considering the cost-effectiveness of attacking a component, and further provided the attack centrality based greedy  algorithms and optimal  algorithms, which embody the framework of PSO. The performance of these algorithms have been validated through the experiments on IEEE bus data. A promising future direction would be to develop a multiple-object optimization framework for the cost-restrained attack problem. Our work helps in assessing the  vulnerability of power grid infrastructures.
%tell that when  attack cost is concerned, which is usually the case in practice,  we need to  make trade-off between effectiveness and cost in the attacks.  different types of compoents. needs to consider the balance among contrict metrics, and condidr cost effecitiveenss in measure the centrality, design algorithms.

% studied the cost restrained hybrid attacks which considers attacking nodes and links simultaneously in power grids. We formulated an optimization problem and showed that the problem is NP-hard in general. Moreover, we proposed a local attack centrality measurement and a greedy hybrid attack algorithm according to the cost performance of attacking a component. We further proposed an improved optimization solution based on the particle swarm optimization framework. The simulation results show that our proposed local attack centrality measurement is effective, and the proposed optimization solution outperforms the considered benchmark methods.

\ifCLASSOPTIONcaptionsoff
  \newpage
\fi

%\begin{thebibliography}{30}
%\bibliographystyle{IEEEtran}
%\bibliography{IEEEabrv,mylib}
%\end{thebibliography}

\end{document}